\documentclass[10pt, conference, letterpaper]{IEEEtran}

\usepackage{amsmath}
\usepackage{amsthm}
\usepackage[ruled, noline, linesnumbered, commentsnumbered, noend]{algorithm2e}
\usepackage{thm-restate}
\usepackage[end]{algpseudocode}
\usepackage{xspace}
\usepackage[table,usenames,dvipsnames]{xcolor}
\usepackage[symbol]{footmisc}
\usepackage{booktabs, tabularx, makecell, multirow}
\usepackage{thmtools}
\usepackage[shortlabels]{enumitem}
\usepackage{graphicx}
\usepackage{subcaption}
\usepackage{cite}
\usepackage{hyperref}
\hypersetup{
 colorlinks,
 citecolor = NavyBlue,
 urlcolor  = Blue,
 linkcolor = ForestGreen,
}

\def\BibTeX{{\rm B\kern-.05em{\sc i\kern-.025em b}\kern-.08em
    T\kern-.1667em\lower.7ex\hbox{E}\kern-.125emX}}
\definecolor{arashcolor}{rgb}{0.0735, 0.49, 0.98}

\newcommand{\system}{\emph{SeedTree}\xspace}

\providecommand{\myRank}[1]{\lfloor \log{\lceil \frac{#1}{c} \rceil} \rfloor}
\providecommand{\wsRank}[1]{\lfloor \log{\lceil \frac{\rank_{t}(#1)}{c} \rceil} \rfloor}
\providecommand{\myLevel}[1]{\lfloor \log{\lceil \frac{#1}{c} \rceil} \rfloor}
\providecommand{\myFloor}[1]{\lfloor #1 \rfloor}

\DeclareMathOperator{\rank}{rank}
\DeclareMathOperator{\level}{level}
\DeclareMathOperator{\inv}{inv}

\newtheorem{definition}{Definition}
\newtheorem{myLemma}{Lemma}
\newtheorem{myTheorem}{Theorem}
\newtheorem{myObservation}{Observation}

\newcommand{\node}{node\xspace}

\newcommand{\nodes}{nodes\xspace}
\newcommand{\Nodes}{Nodes\xspace}

\title{SeedTree: A Dynamically Optimal and \\ Local Self-Adjusting Tree
\thanks{This project has received funding from the European Research Council (ERC) under grant agreement No.\ 864228 (AdjustNet), 2020-2025.}
}

\author{\IEEEauthorblockN{Arash Pourdamghani\textcolor[HTML]{c50e1f}{$^1$}, Chen Avin\textcolor[HTML]{f5951b}{$^2$}, Robert Sama\textcolor[HTML]{04649a
}{$^3$}, Stefan Schmid\textcolor[HTML]{c50e1f}{$^1$}$^,$\textcolor[HTML]{019879}{$^4$}}
\IEEEauthorblockA{\textcolor[HTML]{c50e1f}{$^1$}\textit{TU Berlin, Germany}
\textcolor[HTML]{f5951b}{$^2$}\textit{School of Electrical and Computer Engineering,
Ben Gurion University of the Negev, Israel}
\\
\textcolor[HTML]{04649a
}{$^3$}\textit{Faculty of Computer Science, University of Vienna, Austria}
\textcolor[HTML]{019879}{$^4$}\textit{Fraunhofer SIT, Germany}
}
}

\IEEEoverridecommandlockouts
\begin{document}
\bstctlcite{IEEEexample:BSTcontrol}
\maketitle

\begin{abstract}
We consider the fundamental problem of designing a self-adjusting tree, which efficiently and locally adapts itself towards the demand it serves (namely accesses to the items stored by the tree nodes), striking a balance between the benefits of such adjustments (enabling faster access) and their costs (reconfigurations). This problem finds applications, among others, in the context of emerging demand-aware and reconfigurable datacenter networks and features connections to self-adjusting data structures. Our main contribution is \system, a dynamically optimal self-adjusting tree which supports local (i.e., greedy) routing, which is particularly attractive under highly dynamic demands.
\system relies on an innovative approach which defines a set of unique paths based on randomized item addresses,
and uses a small constant number of items per node. We complement our analytical results by showing the benefits of \system empirically, evaluating it on various synthetic and real-world communication traces.
\end{abstract}


\section{Introduction}
\label{sec: intro}
This paper considers the fundamental problem of designing self-adjusting trees:
trees which adapt themselves towards the demand they serve.
Such self-adjusting trees need to strike an efficient tradeoff between
the benefits of such adjustments (better performance in the future) and their costs
(reconfiguration overheads now). 
The problem is motivated by the fact that workloads in practice 
often feature much temporal and spatial structure, which may be exploited by self-adjusting optimizations~\cite{sigmetrics20complexity,Benson2010}.
Furthermore, such adjustments are increasingly available, as researchers and practitioners are currently making great efforts to render networked and distributed systems more flexible, supporting 
dynamic reconfigurations, e.g., by leveraging programmability (via software-defined networks)~\cite{csur21,pieee19}, network virtualization~\cite{FischerBBMH13}, or reconfigurable optical communication technologies~\cite{osn21}. 

In particular, we study the following abstract model (applications will follow):  
we consider a binary tree 
which serves access requests, issued at the root of the tree, to the items stored by the nodes. 
Each node (e.g., server) stores up to $c$ items (e.g., virtual machines), where $c$ is a parameter indicating the \emph{capacity} of a node. 
We consider an online perspective where items are requested over time. 
An online algorithm aims to optimize the tree in order to minimize the cost of future access requests (defined as the path length between root and accessed item), while minimizing the number of items moving up or down in the tree: the \emph{reconfigurations}. 
We call each movement a reconfiguration, and keep track of its cost.
In particular, the online algorithm which does not know the future access requests, aims to be competitive with an optimal offline
algorithm that knows the entire request sequence ahead of time. In other words,  we are interested in an online algorithm with
minimum \emph{competitive ratio}~\cite{borodin2005online} over any (even worst-case) request sequence.

Self-adjusting trees are not only one of the most fundamental topological structures of their own merit, they also have interesting applications. For example, such trees are a crucial building block for more general
self-adjusting networks: Avin et al.~\cite{avin2017demand} 
recently showed that multiple
trees optimized individually for a single root, 
can be combined to build general communication networks which
provide low degree and low distortion.
The design of a competitive self-adjusting tree
as studied in this paper, is hence a stepping stone.

Self-adjusting trees also feature interesting connections to 
self-adjusting data structures (see \S\ref{sec: related work} for a detailed discussion), for some of which designing and proving
constant-competitive online algorithms is still an open question~\cite{splaytrees}.
Interestingly, a recent result shows that constant-competitive online algorithms exist for self-adjusting balanced binary trees if one maintains a \emph{global} map of the items in the tree; it was proposed to store such a map centrally, 
at a logical root~\cite{ton22push}.
In this paper, we are interested in the question whether this limitation can be overcome,
and whether a competitive \emph{decentralized} solution exist. 

Our main contribution is a dynamically optimal self-adjusting tree, \system
\footnote[1]{The name is due to the additional capacity in nodes of the tree, which resembles seeds in fruits of a tree.}, which achieves a constant competitive ratio
by keeping recently accessed items closer to the root, ensuring a working set theorem~\cite{splaytrees}. 
Our result also implies weaker notions such as key independent optimality~\cite{iacono2005key} (details will follow). 
\system further supports \emph{local} (that is, greedy and hence decentralized) routing,
which is particularly attractive in dynamic networks, by relying on an innovative and simple routing 
approach that enables nodes to take local forwarding decisions: 
\system hashes items to \emph{i.i.d.}~random addresses and defines a set of greedy paths based on these addresses. 
A main insight from our work is that a constant competitive ratio with locality property can be achieved if nodes feature small constant \emph{capacities}, that is, by allowing nodes to store a small constant number of items.
Storing more than a single item on a node is often practical, e.g., on a server or a peer~\cite{ConsistentChord03}, and it is common in hashing data structures with collision~\cite{MirrokniTZ18,AamandKT21}.
We also evaluate \system empirically, both on synthetic traces with ranging temporal locality and also data derived from Facebook datacenter networks~\cite{sigmetrics20complexity}, 
showing how tuning parameters of the \system can lower the total (and access) cost for various scenarios. 

The remainder of the paper is organized as follows. \S\ref{sec: model} introduces our model and preliminaries. 
We present and analyze 
our online algorithm 
in~\S\ref{sec: online}, and transform it to the matching model of datacenter networks in~\S\ref{sec: Matching}.
After discussing our empirical evaluation results in~\S\ref{sec: evaluation}, we review related works in~\S\ref{sec: related work} and conclude our contributions in~\S\ref{sec: conclusion}.

\begin{figure}[t]
    \centering
    \includegraphics[ width=0.32\textwidth]{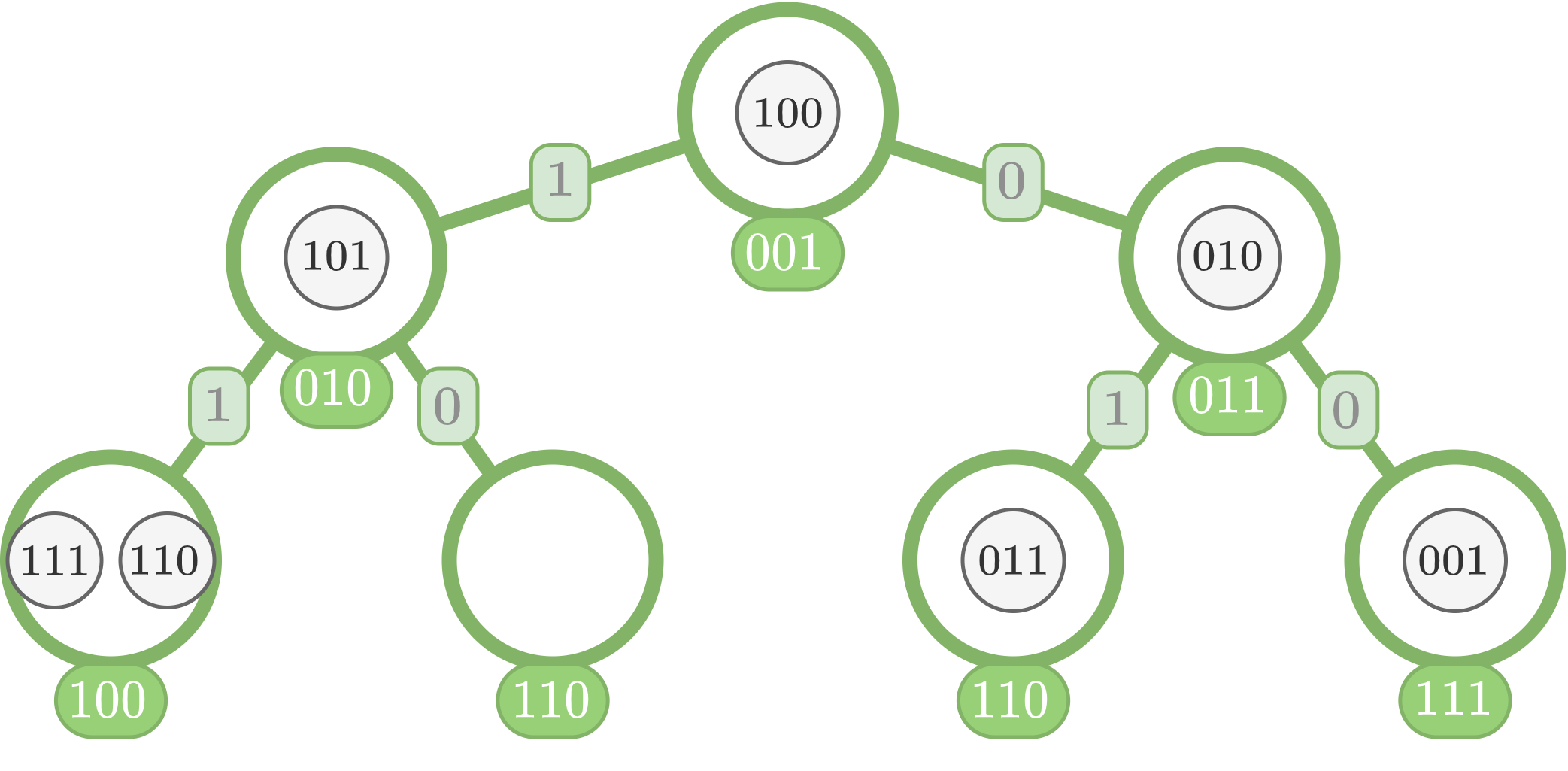}
    \caption{A depiction of \system with capacity 2. Large circles represent \nodes (nodes) of the system, and small circles represent items. The number inside each small circle is the hash of the corresponding item.}
    \label{fig: Intro}  
\end{figure}

\section{Model and Preliminaries}
\label{sec: model}

This section presents our model and introduces preliminaries used in the design of \system. 

\textbf{Items and \nodes.}
We assume a set of \emph{items} $V = (v_1,\dots,v_n)$, and a set of \emph{\nodes} $S = (s_1,\dots)$\footnote[2]{We assume the set of \nodes to be arbitrarily large, as the exact number of \nodes will be determined based on their used capacity.}
arranged as a binary tree $T$. We call the \node $s_1$ the root, which is at \emph{depth} $0$ in the tree $T$, and a \node $s_j$ is at depth $\myFloor{\log{j}}$.

Each \node can store $c$ items, where $c$ is a parameter indicating the \emph{capacity} of a \node.
In our model, we assume that $c$ is a constant. The assignment of items to \nodes can change over time.
We say a \node is \emph{full} if it contains $c$ items, and \emph{empty} if it contains no item 
(See an example in Figure~\ref{fig: Intro}).

We define the level of item $v$ at time $t$,  $\level_t(v)$, as the depth of the \node containing $v$. 
For example, if item $v$ is at \node $s_j$ at time $t$, 
we have $\level_t(v)=\myFloor{\log{j}}$.

\textbf{Request Sequence and Working Set.}
Items are requested over time in an online manner, modeled as a \emph{request sequence} $\sigma = (\sigma_1, \dots, \sigma_m)$, where $\sigma_t = v \in V$ means item $v$ is requested at time~$t$.
We are sometimes interested in the recency of item requests, particularly the size of the working set. Formally, we define $ws_t(\sigma,v)$ as the \emph{working set} of item $v$ in at time $t$ in the request sequence $\sigma$. The working set $ws_t(\sigma,v)$ is a set of unique items requested since the last request to the item $v$ before time $t$.
We define a \emph{rank} of item $v$ at time $t$, $\rank_t(v)$, as the size of working set of the item $v$ at time $t$.


\textbf{Costs and Competitive Ratio.}
We partition costs incurred by an algorithm, $ALG$, into two parts, the cost of finding an item: the \emph{access cost}, and the cost of reconfigurations: the \emph{reconfiguration cost}.
The search for any item starts at the root \node and ends at the \node containing the item. 
Based on our assumption of constant capacity,
we assume the cost of search inside a \node to be negligible.
Furthermore, assuming the local routing property, we find an item by traversing a single path in our tree; hence the access cost for an access request $\sigma_i$, $C_{ALG}^A(\sigma_i)$, equals the level at which the item is stored.

In our model, a reconfiguration consists of moving an item one level up or one level down in the tree, plus potentially additional lookups inside a \node.
We denote the total reconfiguration cost after an access request $\sigma_i$ by $C_{ALG}^R(\sigma_i)$.
Hence, the total cost of each access request
is $C_{ALG}^A(\sigma_i) + C_{ALG}^R(\sigma_i)$, and the total cost of the algorithm on the whole request sequence is:
$C_{ALG}(\sigma) = \sum_{i=1}^m C_{ALG}^A(\sigma_i) + C_{ALG}^R(\sigma_i)$.
The objective of \system is to operate at the lowest possible cost, or more specifically, as close as possible to the cost of an optimal offline algorithm, $OPT$.

\begin{definition}[Competitive ratio] Given an online algorithm $ALG$ and an optimal offline algorithm $OPT$, the (strict) competitive ratio is defined as:
$
    \rho_{ALG} = \max_{\sigma} \frac{C_{ALG}(\sigma)}{C_{OPT}(\sigma)}
$
\end{definition}

Furthermore, we say an algorithm has (strict) \emph{access competitive ratio}  considering only the access cost of the online algorithm $ALG$ (not including the reconfiguration cost).

In this paper, we prove that \system is \emph{dynamically optimal}. It means that the cost of our algorithm matches the cost of the optimal offline algorithm asymptotically.

\begin{definition}[Dynamic optimality]
Algorithm $ALG$ is dynamically optimal if it has constant competitive ratio, i.e., $\rho_{ALG} = O(1)$.
\end{definition}

\textbf{$MRU$ trees.}
We define a specific class of self-adjusting trees, $MRU$ trees.
An algorithm maintains a $MRU$ tree if it keeps items at a similar level to their ranks. 

\begin{figure*}[t]
    \begin{subfigure}[b]{0.33\linewidth}
    \centering
    \includegraphics[width=\textwidth]{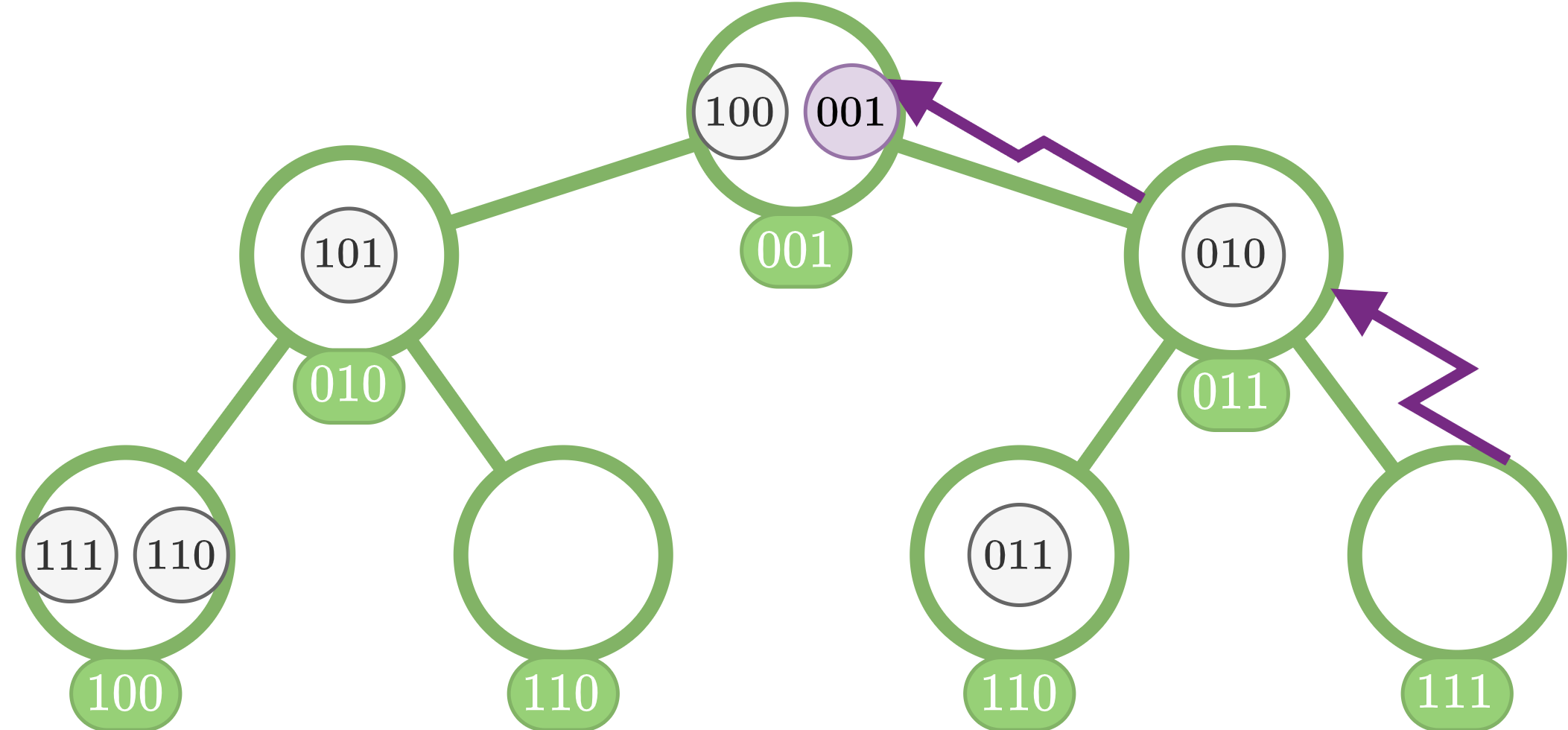}
    \caption{ 
    Item $001$ moves up, \node-by-\node, until it reaches the root.}
    \label{subfig: move-to-root}
    \end{subfigure}
     \begin{subfigure}[b]{0.33\linewidth}
    \centering
    \includegraphics[width=\textwidth]{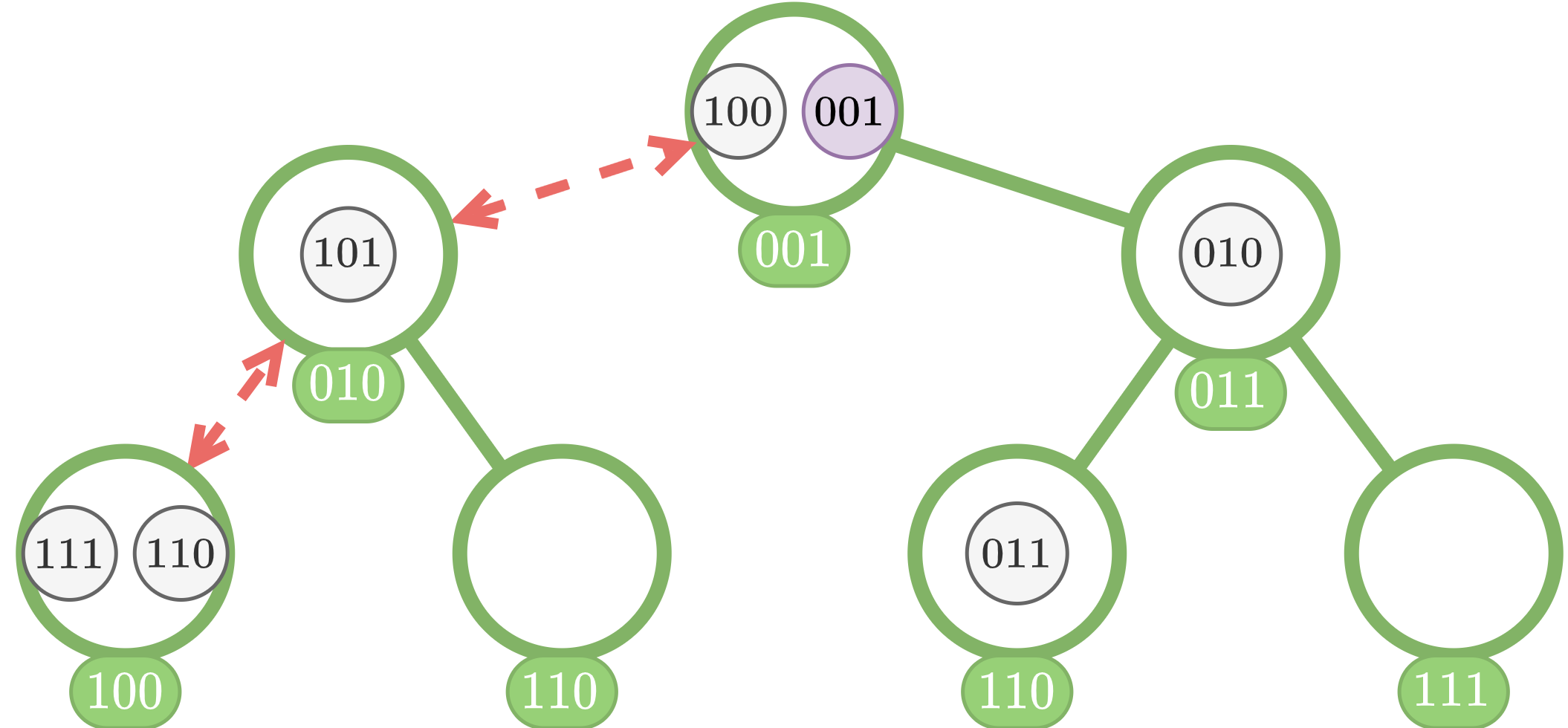}
    \caption{The first try of push-down failed, because \node $100$ is full.}
    \label{subfig: failed}
    \end{subfigure}
      \begin{subfigure}[b]{0.33\linewidth}
    \centering
    \includegraphics[width=\textwidth]{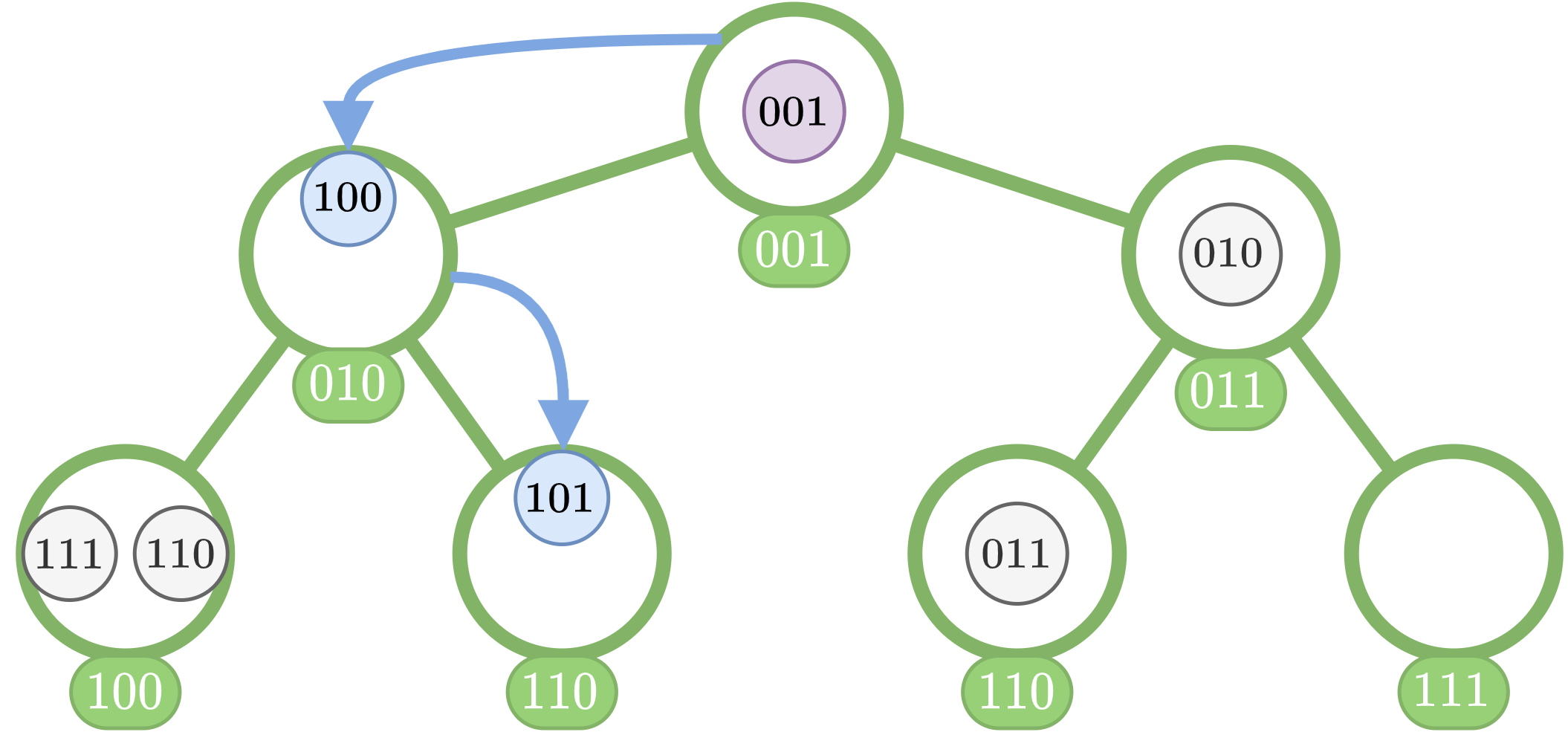}
    \caption{After finding non-full \node, items are pushed down \node-by-\node.}
    \label{subfig: push-down}
    \end{subfigure}
    \caption{
    An example of steps taken in Algorithm~\ref{Alg: online}, starting from the state of \system in Figure~\ref{fig: Intro}, which has a capacity equal to $2$. In this example, the request is an access request to the item with the hash value $001$ (the purple circle).
    Subfigure~\ref{subfig: move-to-root} shows the move-to-the-root phase, and Subfigures~\ref{subfig: failed} and~\ref{subfig: push-down} depict the push-down phase.
    }
    \label{fig: online}  
\end{figure*}

\begin{definition}[$MRU$ tree]
An algorithm has the $MRU(0)$ property if for any item $v$ inside its tree and at any given time $t$, the equality $\level_t(v) = \myRank{\rank_t(v)}$ holds. 

Similarly, we say an algorithm maintains an $MRU(\beta)$ if it ensures the relaxed bound of $\level_t(v) \le \myRank{\rank_t(v)} + \beta$ for any item $v$ in the tree.
\end{definition}

\section{Online SeedTree}
\label{sec: online}

This section presents \system, an online algorithm that is dynamically optimal in expectation. This algorithm build upon uniformly random generated addresses, and allows for local routing, while ensuring dynamic optimality.
Details of the algorithm are as follows:
Algorithm~\ref{Alg: online} always starts from the root \node. Upon receiving an access request to an item $v$
it performs a local routing (described in Procedure~\ref{Alg: localSearhch}) based on the uniformly random binary address generated for the node $v$, which uniquely determines the path of $v$ in the tree.
We call the $i$-th bit of the address of $v$ by $H(v,i)$.
Let us assume that the local routing for node $v$ ends in level~$\ell$.
\begin{procedure}
  \small
  \caption{LocalRouting($s$,$v$)}
  \label{Alg: localSearhch}
            \If{$H(v,\level(s))$ equals 0}{
                Return the left child of $s$.
            }
            \Else{
                Return the right child of $s$.
            }
\end{procedure}

Then \system performs the following two-phase reconfiguration. These two phases are designed to ensure the level of items remains in the same range as their rank (details will follow), and the number of items remains the same at each level.
\begin{enumerate}
   \item \emph{Move-to-the-root:}
    This phase moves the accessed item to the \node at the lowest level possible, the root of the tree.
    The movement of the item is step-by-step, and it keeps all the other items in their previous \node (we keep the item in a temporary buffer if a \node on the path was full).
    This phase is depicted in Figure~\ref{subfig: move-to-root} by zig-zagged purple arrows.
    \item \emph{Push-down:} 
    In this phase, our algorithm starts from the root \node,  selects an item in the \node (including the item that has just moved to this \node) uniformly at random, and moves this item one level down to the new \node selected in the~\ref{Alg: localSearhch} procedure.
    The same procedure is continued for the new \node until reaching level $\ell$, the level of the accessed item.
    If the \node at level $\ell$ was non-full, the re-establishment of balance was successful.
    Otherwise, if this attempt is failed, the algorithm reverses the previous push downs back to the root, and starts again, until an attempt is successful.
    As an example, the failed attempt of this phase is depicted by dashed red edges in Figure~\ref{subfig: failed} and the last successful one by curved blue arrows in Figure~\ref{subfig: push-down}.
\end{enumerate}

\begin{algorithm}[h]
\caption{Online \system}
\label{Alg: online}
    \small
    \SetKwInOut{KwIn}{Input}
    \KwIn{Accessed item $v$.}
    Set $s$ as the root.\\
    \While{$s$ does not contain $v$}{
        $s$ = \ref{Alg: localSearhch}($s$,$v$).
    }
    Call the current level of $v$ as $\ell$.\\
    Set $s$ as the root, and move item $v$ to $s$.\\
    \label{line: Move up}
    \While{balance is not fixed}{
        Call the current \node $s$.\\
        \While{level of $s$ is less than $\ell$}{
        \label{line: Start Move down} 
        Take an item in \node $s$, uniformly at random, call it $v$.\\
        \label{line: random} 
        $s$ = \ref{Alg: localSearhch}($s$,$v$).\\
        Add item $v$ to the \node $s$.
        }
        \If{the last chosen \node is full}{
            Reverse the push-down back to the root.
        }
    }
    \label{line: End Move down}
\end{algorithm}

Algorithm~\ref{Alg: online} always terminates, as there is always the chance that the item which has been moved to root is selected among all candidates, and we know that the \node which that item is taken from is not full.
We now state the main theorem of the paper that proves the dynamic optimality of \system. 
\begin{myTheorem}
\label{thm: online-algorithm}
SeedTree is dynamically optimal for any given capacity $c \geq 1$.
\end{myTheorem}
The proof of Theorem~\ref{thm: online-algorithm} is at the end of the section. The first step towards the proof is showing that the number of items in each level remains the same. It is true because after removing an item at a certain level, the algorithm adds an item to the same level as a result of the push-down phase.
\begin{myObservation}
\label{obs: consistent}
SeedTree keeps the number of items the same at each level.
\end{myObservation}

The rest of the analysis is based on the assumption that the algorithm was initialized with a fixed \emph{fractional occupancy} $0<f<1$ of the capacity of each level, i.e., in level $i$, the initial tree has exactly $\lfloor c \cdot f \cdot 2^i \rfloor$ items. At the end of this section, we will see that $f = \frac{1}{2}$ works best for our analysis. However, we emphasize that having $0<f<1$ suffices for \system to run properly.

The second observation is a result of Observation~\ref{obs: consistent}. As the number of items remains the same in each level (based on Observation~\ref{obs: consistent}) at most a fraction $f$ of all \nodes are full. In the lowest level, the number of full \nodes might be even lower; hence the probability of a uniformly random \node being full is at most $f$ when we go to the next request.

\begin{myObservation}
\label{obs: full}
Algorithm~\ref{Alg: online} ensures that the probability of any uniformly random chosen \node in SeedTree to be full, after serving each access request, is at most $f$.
\end{myObservation}

According to Algorithm~\ref{Alg: online}, items are selected uniformly at random inside a \node. In the following lemma, we show that a \node in a certain level is also selected uniformly at random, which enables the rest of the proof.

\begin{myLemma}
\label{lem: uniform}
\Nodes selected on the final path of the push-down phase with a level lower than $\ell$ are selected uniformly at random.
\end{myLemma}
\begin{proof}
Let us denote the probability of $\ell'$-th \node on the path (the \node at level $\ell'$, denoted by $s_{\ell'}$) being the selected \node is $\frac{1}{2^{\ell'}}$. Our proof goes by induction. For the basis, $\ell'=0$, it is true since we only have one \node, the root.
Now assume that in the final path of push down, we want to see the probability of reaching the current \node, $s_{\ell'}$. Based on the induction assumption, we know that the parent of $s_{\ell'}$, the \node $s_{\ell'-1}$, has been selected uniformly at random, with probability $\frac{1}{2^{\ell'-1}}$. Based on Line~\ref{line: random} of Algorithm~\ref{Alg: online}, an item is selected from those inside $s_{\ell'-1}$ uniformly at random, plus having  the independence guarantee of our hash function that generated address of the selected item, we can conclude the decision to go to left or right from $s_{\ell'-1}$ was also uniformly at random, hence the probability of reach $s_{\ell'}$ is $\frac{1}{2^{\ell'-1}} \cdot \frac{1}{2} = \frac{1}{2^{\ell'}}$.
 Note that the above-mentioned choices are independent of whether or not the descents $s_{\ell'-1}$  are full or not. Hence the choice is independent of (possible) previous failed attempts of the push-down phase (which might happen due to having a full \node at level $\ell$), i.e., the previous attempts do not affect the probability of choosing the \node $s_{\ell'}$.
\end{proof}

An essential element of the proof of Theorem~\ref{thm: online-algorithm} is that the rank and level of items are related to each other. Lemma~\ref{lem: depth} describes one of the aspects of this relation.

\begin{myLemma}
\label{lem: depth}
During the execution of the SeedTree, for items $v$ and $u$ at time $t$, if $\rank_t(v) > \rank_t(u)$ then $E[\level_t(v)] > E[\level_t(u)]$.
\end{myLemma}
\begin{proof}

Having $\rank_t(v) > \rank_t(u)$, we know that $u$ was accessed more recently than $v$. Let us consider time $t'$, the last time $u$ was accessed. Since the rank of $v$ is strictly larger than the rank of $u$, and as $u$ was moved to the root at time $t'$,  we know that  $\level_{t'}(v)> \level_{t'}(u)$. 

Items $u$ and $v$ might reach the same level after time $t'$, but it is not a must. We consider the level that they first met as a random variable, $L_{uv}$. We denote $L_{uv} = -1$ if $u$ and $v$ never appear on the same level after time $t'$.
Let us quantify the difference in the expected level of $u$ and $v$, using the law of total expectation:
\[E[\level_t(v)]-E[\level_t(u)]\]
\[ =\sum_{k=-1}^{\myLevel{n}} Pr(L_{uv} = k) \cdot(E[\level_t(v)| L_{uv} = k]
\]\[- E[\level_t(u)| L_{uv} = k])
\]

For the case $L_{uv} =-1$, we know that $u$ and $v$ never reached the same level, and the following is always true: 
\[E[\level_t(v)| L_{v,u} = -1]>E[\level_t(u)| L_{v,u} = -1]\]

For $k \ge 0$, let us consider the time $t''$ when $u$ and $v$ meet at the same level, i.e $\level_{t''}(u) = \level_{t''}(v)$. After items $u$ and $v$ meet for the first time, their expected progress is the same. More precisely, consider the current subtree of the \node containing $v$ at time $t''$, and call it $T'$. Since the item addresses are chosen uniformly at random, the expected number of times that $T'$ is a subtree of a \node containing $v$, equals the number of times that $T'$ might be a subtree of \node containing $u$ in the same level. Hence the expected increase in the level for both items $u$ and $v$ stays the same from time $t''$ onward.
\end{proof}

Next, we explain why the number of items accessed at a higher level is limited in expectation for any given item.

\begin{restatable}{myLemma}{higherLevel}
\label{Lem: higher}
For a given item $v$ at time $t$, there are at most $2 \cdot \rank_t^{ws}(v)$ items accessed at a higher level since the last time $v$ was accessed, in expectation. 
\end{restatable}
\begin{proof}
Given Lemma~\ref{lem: depth}, the proof is along the lines of the proof of Lemma~4 from~\cite{ton22push}. We removed the details of the proof due to space constraints.
\end{proof}

Now we prove the items in the tree maintained by the online \system are not placed much farther from their position in a tree that realizes the exact working set property. This in turn allows us to approximate the total cost of the online \system in comparison to the optimal offline algorithm with the same capacity. The approximation factor, $2-\log(f)$, is intuitive: with less capacity in each level (lower values of levels' fractional occupancy), we need to put items further down.

\begin{myLemma}
\label{lem: ALG-MRU}
SeedTree is $MRU(2-\log(f))$ in expectation.
\end{myLemma}
\begin{proof}
For any given item $v$ and time $t$, we show that $E[\level_t(v)] \le \wsRank{v} + 2-\log(f)$ remains true, considering move-to-the-root and push-down phases.
As can be seen in Line~\ref{line: Start Move down} of Algorithm~\ref{Alg: online}, the item $v$ might move down if the current level of $v$ is lower than the level of the accessed item.

Let us denote the increase in the level from time $t'$ to time $t$ by a random variable $D(t',t)$.
We express this increase in terms of an indicator random variable $I(t',t,\ell)$ which denotes whether item $v$ went down from level $\ell$ during $[t',t]$ or not. We know that:
\[D(t',t) = \sum_{\ell} I(t',t,\ell)\]

Let $K$ denote the number of items accessed from a higher level, and let us write $K = k_1+\dots+k_{\lceil \frac{n}{c} \rceil}$, where  $k_{\ell}$ means that $k_{\ell}$ such accesses happened when item $v$ was at level $\ell$.
For the level $\ell$, based on the Observation~\ref{obs: consistent} and Lemma~\ref{lem: uniform} and the fact that each level contains $f \cdot c \cdot 2^{\ell}$ items, we conclude $v$ is being selected after $k_{\ell}-1$ accesses with probability $(1- \frac{1}{f \cdot c \cdot 2^{\ell}})^{k-1} \cdot (\frac{1}{f \cdot c \cdot 2^{\ell}})$.

\[I(t',t,\ell) = \min(1, \sum_{k_{\ell} = 0}^{K} (1- \frac{1}{f \cdot c \cdot 2^{\ell}})^{k_{\ell}-1} \cdot (\frac{1}{f \cdot c \cdot 2^{\ell}}) )
\]\[= \min( 1, (\frac{1}{f \cdot c \cdot 2^{\ell}}) \cdot \sum_{k_{\ell} = 0}^{K} (1- \frac{1}{f \cdot c \cdot 2^{\ell}})^{k_{\ell}-1} )
\]
\[
\le \min(1,(\frac{K}{f \cdot c \cdot 2^{\ell}}))
\]

Going back to our original goal of finding how many levels an item goes down during a time period $[t',t]$, we have:
\[E[D(t',t)] \le \sum_{\ell} E[\min(1,(\frac{K}{f \cdot c \cdot 2^{\ell}}))] 
\]\[ = \log(\frac{E[K]}{f \cdot c}) + 1 = \log(E[K])-
\log(c) - \log(f) + 1 
\]

The last equality comes from the fact that for $\ell = \log(\frac{E[K]}{f \cdot c})$, we have $\frac{K}{f \cdot c \cdot 2^{\ell}} \le  1$, and for all larger values of $\ell$, the value will decrease exponentially with factors of two.

From Lemma~\ref{Lem: higher} we know that the expected value of $K$ is less than equal to $2 \cdot \rank_t(v)$; therefore, the expected increase is: 
\[E[D(t',t)] \le \log(2 \cdot \rank_t(v)) - \log(c) - \log(f) + 1 \]\[= \log(\rank_t(v)) - \log(c) + 2 - \log(f) 
\]
\end{proof}

The following lemma shows the relation between the total cost of the online \system and fractional occupancy $f$. The relation is natural: as $f$ becomes smaller, the chance of finding a non-full \node becomes larger, and thus fewer attempts are needed to find a non-full \node.

\begin{myLemma}
\label{lem: Reconfigure}
The expected cost of SeedTree is less than equal to $2\cdot(\lceil \frac{1}{(1-f)} \rceil + 1)$ times the access cost.
\end{myLemma}
\begin{proof}
Let us consider the accessed item $v$ at level $\ell$. 
In the first part of the algorithm, the move-to-the-root phase costs the same as the access, which is equal to traversing $\ell$ edges.
As the probability of a \node being non-full is $1-f$ based on Observation~\ref{obs: full}, and as the choice of \nodes is uniform based on Observation~\ref{lem: uniform}, only $\lceil \frac{1}{1-f} \rceil$ iterations are needed during the push-down phase for finding a non-full \node, each at cost $2 \cdot \ell$.
Hence, given the linearity of expectation, we have:
\[ E[C_{ALG}] = E[C^{\text{Access}}_{ALG}+C^{\text{Move-to-the-root}}_{ALG} + C^{\text{Push-down}}_{ALG}] 
\]
\[\le 2 \cdot (1+\lceil \frac{1}{1-f} \rceil) \cdot \ell = 2 \cdot (1+\lceil \frac{1}{1-f} \rceil) \cdot C^{\text{Access}}_{ALG} \]
\end{proof}

We now describe why working set optimality is enough for dynamic optimality, given that reconfigurations do not cost much (which is proved in Lemma~\ref{lem: Reconfigure}). Hence, any other form of optimality, such as key independent optimality or finger optimality is guaranteed automatically~\cite{iacono2005key}.

\begin{restatable}{myLemma}{MRUC}
\label{lem: MRUC}
For any given $c$, an $MRU(0)$ algorithm is $(1+e)$ access competitive. 
\end{restatable}
\begin{proof}
The proof relies on the potential function argument. 
We describe a potential function at time $t$ by $\phi_t$, and show that the change in the potential from time $t$ to $t+1$ is $\Delta \phi_{t \rightarrow t+1}$. 

Our potential function at time $t$, counts the number of items that are misplaced in the tree of the optimal offline algorithm $OPT$ with regard to their rank. 
(As the definition of $MRU(0)$ indicates, there exists no inversion in such a tree, that is why we only focus on the number of inversions in $OPT$.)
Concretely, we say a pair $(v,u)$ is an inversion if $\rank_t(v)<\rank_t(u)$ but $\level_t(v) > \level_t(u)$.
We denote the number of items that have an inversion with item $v$ at time $t$ by $\inv_t(v)$, and define $B_t(v) = 1 + \frac{\inv_t(v)}{c\cdot 2^{\level_t(v)}}$. 
Furthermore, define $B_t = \prod_{v=1}^n B_t(v)$. 
We define the potential function at time $t$ as $\phi_t = \log B_t$.
We assume that the online \system rearranges its required items in the tree before the optimal algorithm's rearrangements. 
Let us first describe the change in potential due to rearrangement in the online \system after accessing item $\sigma_t = v$. This change has the following effects:
\begin{enumerate}
    \item Rank of the accessed item, $v$, has been set to $1$.
    \item Rank of other items in the tree might have been increased by at~most~$1$.
\end{enumerate}

Since the relative rank of items other than $v$ does not change because of the second effect, it does not affect the number of inversions and hence the potential function.
Therefore, we focus on the first effect. Since $OPT$ has not changed its configuration, for all items $u$ that  are being stored in a lower level than $v$ in the $OPT$, a single inversion is created, therefore we have $B_{t+1}(u) = B_{t}(u) + \frac{1}{c \cdot 2^{\level^c(u)}}$.
For the accessed item $v$, as its rank has changed to one, all of its inversions get deleted. The number of inversions for other items, except $v$, remains the same.
Let us denote the number of items with lower level than $v$ at time $t$ by $L_t(v)$ and partition the $\prod_{i=1}^n B_{t+1}(i)$ into three parts as we discussed ($v$, items stored in a lower level than $v$, and other items denoted by set $O_t(v)$):

\[
 \prod_{i=1}^n B_{t+1}(i) =
 B_{t+1}(v) \cdot \prod_{i \in L_t(v)} B_{t+1}(i) \cdot \prod_{i \in O_t(v)} B_{t+1}(i) 
\]
By rewriting $B_{t+1}(i)$ in terms of $B_t(i)$, we get:
\[
 \prod_{i=1}^n B_{t+1}(i) = 1 \cdot  \prod_{i \in L_t(v)} (B_{t}(i)+\frac{1}{c \cdot 2^{\level_t(i)}}) \cdot \prod_{i \in O_t(v)} B_{t}(i) 
\]

Now let us look at potential due the first effect from time $t$ to $t+1$ by $\Delta \phi^1_{t \rightarrow t+1}$, and describe it in more detail:
\[
    \Delta \phi^1_{t \rightarrow t+1} = 
    \log B_{t+1} - \log B_t = \log \frac{B_{t+1}}{B_t}
    \]
    \[
    =\log \frac{\prod\limits_{i=1}^n B_{t+1}(i)}{\prod\limits_{i=1}^n B_t(i)}
 = \log (\frac{1}{B_t(v)} \cdot \frac{\prod\limits_{L_t(v)} (B_t(i)+\frac{1}{c \cdot 2^{\level_t(i)}})}{\prod\limits_{L_t(v)} B_t(i)} )
\]
\[
 \le
 \log (
 \frac{1}{B_t(v)} \cdot e^{|L_t(v)|} )
\]
in which the last inequality comes from the fact that
$|L_t(v)| = c \cdot 2^{\level_t(i)}$ and also the inequality that:
\[
\prod_{i=1}^{|L_t(v)|} (B_t(i)+\frac{1}{|L_t(v)|}) 
\le
\prod_{i=1}^{|L_t(v)|} (B_t(i)+\frac{B_t(i)}{|L_t(v)|})
\]\[=
(1+\frac{1}{|L_t(v)|})^{|L_t(v)|} \cdot \prod_{i=1}^{|L_t(v)|} B_t(i)
\le 
e^{|L_t(v)|} \cdot \prod_{i=1}^{|L_t(v)|} B_t(i)
\]
Now let us focus on $B_t(v)$, and first assume that $\wsRank{v} > \level_t(v)$.
We want to find the maximum number of items that might cause inversion with the accessed item $v$.

Among all $c \cdot 2^{\wsRank{v}}-1$ items that $v$ might have higher rank them, at most $c \cdot 2^{\level_t(v)}-1$ have lower level in the $OPT$ tree. Hence we have:

\[
B_t(v) = \frac { (c \cdot 2^{\wsRank{v}}-1) - (c \cdot 2^{\level_t(v)}-1) }{c \cdot 2^{\level_t(v)}}
\]\[\ge 
\frac { (2^{\wsRank{v}}-1) }{ 2^{\level_t(v)}} - 1 
\]
\[
\ge \frac { 2^{\wsRank{v}} }{ 2^{\level_t(v)+1}} = 2^{\wsRank{v} - \level_t(v)-1}
\]
hence the change in potential due to the first effect is:
\[
 \Delta \phi^1_{t \rightarrow t+1}
 \le 
 \log (\frac{1}{2^{\wsRank{v} - \level_t(v)-1}} \cdot e^{\level_t(v)}) 
 \]\[=
 \log( 2^{(1 + \log e) \cdot \level_t(v) - \wsRank{v}})
 \]
\[
 = 
 (1 + \log e) \cdot \level_t(v) - \wsRank{v}
\]
For the case $\wsRank{v} < \level_t(v)$, we use the fact that $B_v^t > 1$, from the first inequality below:
\[
 \Delta \phi_{t \rightarrow t+1}
 =
 \log (\frac{1}{B_v^t} \cdot e^{\level_t(v)}) 
 \]\[\le
 \log( 2^{\log e \cdot \level_t(v)})
 = 
 \log e \cdot \level_t(v) 
\]
\[
 =(1+\log e) \cdot \level_t(v) - \wsRank{v}
\]
Hence, in both cases of $\wsRank{v}$ being larger or smaller than $\level_t(v)$, we have $\Delta \phi_{t \rightarrow t+1}  \le  (1+\log e) \cdot \level_t(v) - \wsRank{v}$.

We then show changes in the potential because of $OPT$'s reconfiguration. Details of the computations are omitted due to space constraints, but they are similar to the changes in potential due to rearrangements in the $ON$'s algorithm, and the result is that each $OPT$'s movement costs less than $\log e$.

Summing up changes in the potential after $ON$'s and $OPT$'s reconfiguration, assuming $OPT$ has done $w_t$ movements at time $t$, we end up with:
\[\Delta \phi^{t \rightarrow t+1}
=  (1+\log e) \cdot \level_t(v) - \wsRank{v} + w \cdot \log e \]
And hence the cost of the online algorithm $MRU(0)$ at time $t$ is at most:
\[
C^t_{MRU(0)} = 
C^t_{Amortized} + \Delta \phi^{t}
\]
\[
=
\wsRank{v}+ (1+\log e) \cdot \level_t(v) \]
\[- \wsRank{v} + w_t \cdot \log e
\le
(1+\log e) \cdot ( \level_t(v) + w_t)
\]
And then summing up the cost of the $MRU(0)$ and $OPT$ for the whole request sequence, we will get: 
\[
C_{ON} =
 \sum_t C^t_{ON}
 \le \sum_{t} (1+\log e) \cdot ( \level_t(v) + w_t) 
 \]\[
 =
 (1+\log e) \cdot C_{OPT}
\]
In which the last equality comes from the fact that $OPT$ also needs to access the item, and as we assumed an additional $w_t$ reconfigurations.
\end{proof}

As the first application of Lemma~\ref{lem: MRUC} we prove a lower bound on the cost of any online algorithm that only depends on the size of the working set of accessed items in the sequence.
\begin{myTheorem}
\label{thm: lowerbound}
Any online algorithm maintaining a self-adjusting complete binary tree with capacity $c>1$ on a request sequence $\sigma = \sigma_1, \dots \sigma_m$, requires an access cost of at least $\frac{\sum_{i=1}^{m}\myRank{\rank_t(\sigma_i)}}{(1+e)}$.
\end{myTheorem}
\begin{proof}
This proof is an extension and improvement of the proof from~\cite{ton22push} for any values of $c > 2$. A result of Lemma~\ref{lem: MRUC} is that even an optimal algorithm cannot be better than $\frac{1}{(1+e)}$ the $MRU(0)$, otherwise contradicting Lemma~\ref{lem: MRUC}. As the cost of each access to the item $\sigma_i$ is $\myRank{\rank_t(\sigma_i)}$ in $MRU(0)$,  we can conclude the total cost of any algorithm should be larger than $\frac{\sum_{i=1}^{m}\myRank{\rank_t(\sigma_i)}}{(1+e)}$.
\end{proof}

\begin{myLemma}
\label{lem: MRUAB}
Any $MRU(\beta)$ tree is $\beta \cdot (1+e)$-access competitive. 
\end{myLemma}
\begin{proof}
Lemma~\ref{lem: MRUC} shows that an $MRU(0)$ is $(1 + e)$-access competitive.
Any item which was in level $k$ in $MRU(0)$,  is in level $k + \beta$ in $MRU(\beta)$.
As an $MRU(\beta)$ algorithm keeps items with $\rank^c(0)$ at $level(0)$, and because for any $k \ge 1$, we have 
$k + \beta \le \beta k$, 
we obtain that $MRU(\beta)$ is $(\beta) \cdot (1+e) $-access competitive.
\end{proof}

We conclude this section by proving our main theorem, dynamic optimality of online \system.
\begin{proof}[proof of Theorem~\ref{thm: online-algorithm}]
Combining Lemma~\ref{lem: ALG-MRU}, Lemma~\ref{lem: Reconfigure} and Lemma~\ref{lem: MRUAB} yields that
the upper bound for competitiveness is $(1+e) \cdot (2 \cdot (1+\lceil \frac{1}{1-f} \rceil))\cdot (2-\log(f))$.
The fractional occupancy $f=1/2$ in the above formula is the optimal value for $f$, which gives us the $43$-competitive ratio.
\end{proof}
We need to point out that the above calculation is just an \emph{upper bound} on the competitive ratio. As we will discuss in \S\ref{sec: evaluation}, the best results are usually achieved with a slightly higher value of $f$, which we hypothesize might be because of an overestimation of items' depth in our theoretical analysis.

\section{Application in reconfigurable datacenters}
\label{sec: Matching}

\begin{figure}[t]
    \centering
    \includegraphics[ width=0.33\textwidth]{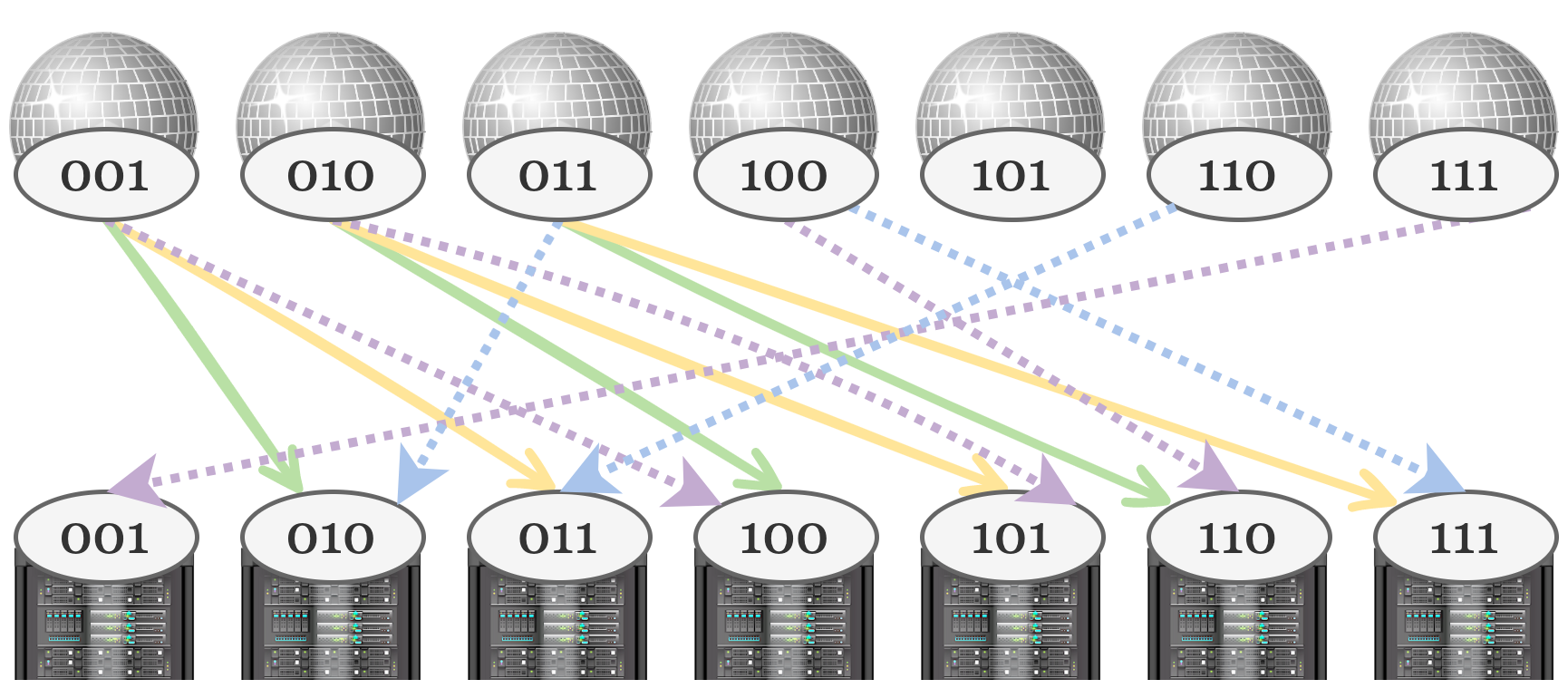}
    \caption{
    A transformation from the example \system shown in Figure~\ref{fig: Intro}, which has capacity $c=2$ and fractional occupancy of $f=\frac{1}{2}$.
    The disco balls on top represent the reconfigurable switches, and below are datacenter racks.
    Solid edges show structural matchings, and dotted edges represent membership matchings.}
    \label{fig: matching}  
\end{figure}

\system provides a fundamental self-adjusting structure which is useful in different settings. 
For example, it may be used to adapt the placement of containers in virtualized settings, in order
to reduce communication costs. However, \system can also be applied in reconfigurable networks
in which links can be adapted. In the following, we describe how to use \system in such a use case
in more detail. In particular, we consider reconfigurable datacenters in which the connectivity
between racks, or more specifically Top-of-the-Rack (ToR) switches, can be adjusted dynamically, e.g., based on optical circuit switches~\cite{osn21}. An optical switch provides a matching between racks,
and accordingly, the model is known as a \emph{matching model} in the literature~\cite{sigmetrics22cerberus}. 
In the following, we will show how a \system with capacity $c$ and fractional occupancy of $f= \frac{1}{c}$ can be seen in terms of $2+c$ matchings, and how reconfigurations can be transformed to the matching model\footnote[3]{The matching model considers perfect matchings only, however, in practice imperfect matchings can be enforced by ignore rules in switches.}. 
We group these matchings into two sets:
\begin{itemize}
    \item \textbf{Topological matchings}: 
    consists of $2$ \emph{static} matchings, embedding the underlying binary tree of \system. The first matching represents edges between a \node and its left child (with the ID twice the ID of the \node), and similarly the second matching for the right children (with the ID twice plus one of the ID of their parents).
    An example is depicted with solid edges in Figure~\ref{fig: matching}.
    \item \textbf{Membership matchings}:
    has $c$ \emph{dynamic} matchings, connecting \nodes to items inside them.
    If a \node has more than one item, the corresponding order of items to matchings is arbitrary.
    An example is shown with dotted edges in Figure~\ref{fig: matching}. 
\end{itemize}

Having the matchings in place, let us briefly discuss how search and reconfiguration operations are implemented.
A search for an item starts at the node with ID $001$, the root node. We then check membership matchings of this node. If they map to the item, we have found the \node which contains the item, and our search was successful. Otherwise, we follow the edge determined by the hash of the item, going to the new possible \node hosting the item. We repeat the process of checking membership matchings and going along topological matchings until we find the item. The item will be found, as it is stored in one of the \nodes in the path determined by its hash value.
Each step of moving an item can be implemented in the matching mode with only one edge removal and one edge addition in membership matchings.

\begin{figure*}[t]
    \begin{subfigure}[b]{0.33\linewidth}
    \centering
    \includegraphics[width=\textwidth]{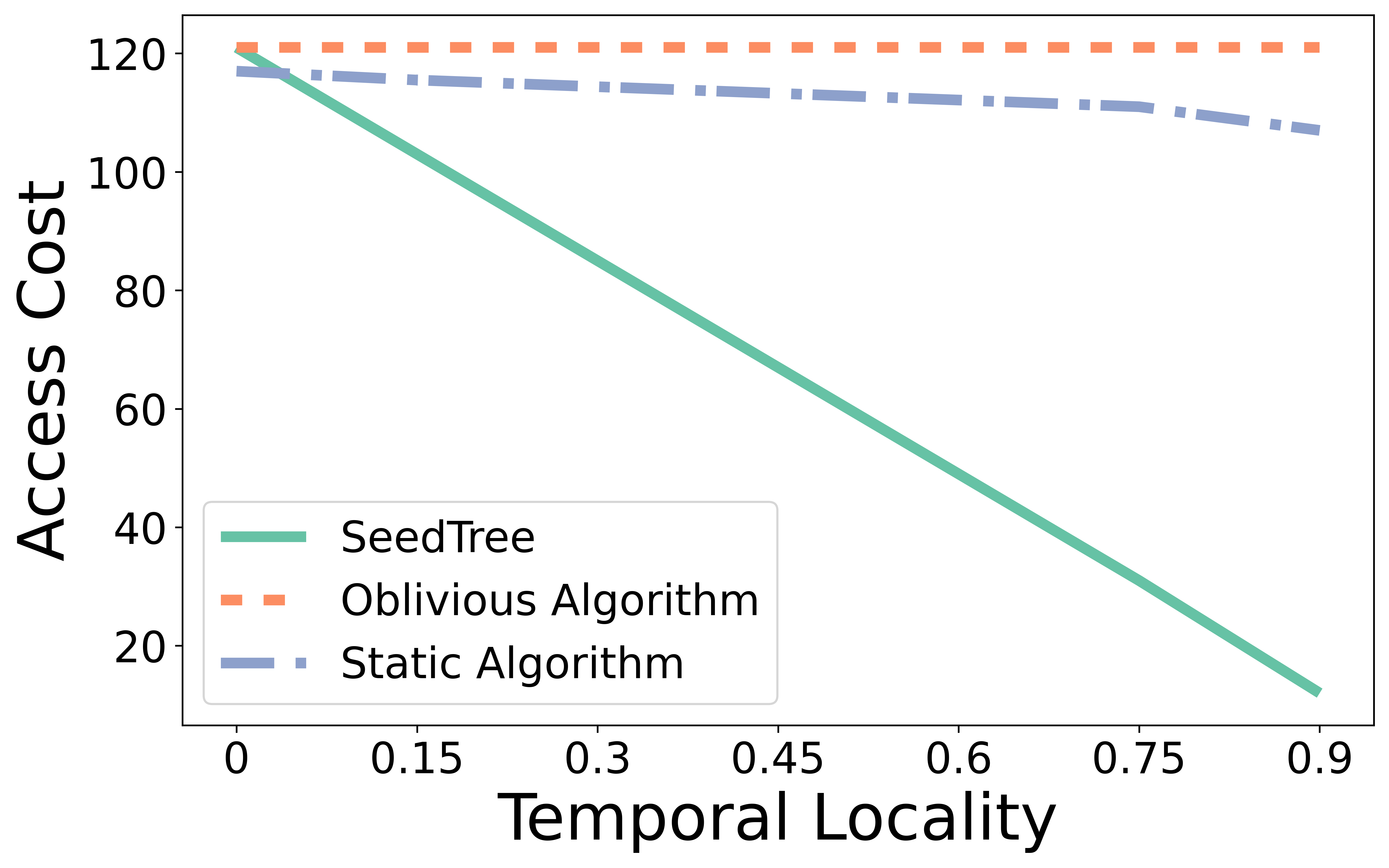}
    \caption{}
    \label{subfig: toObv}  
    \end{subfigure}
    \begin{subfigure}[b]{0.33\linewidth}
    \centering
    \includegraphics[width=\textwidth]{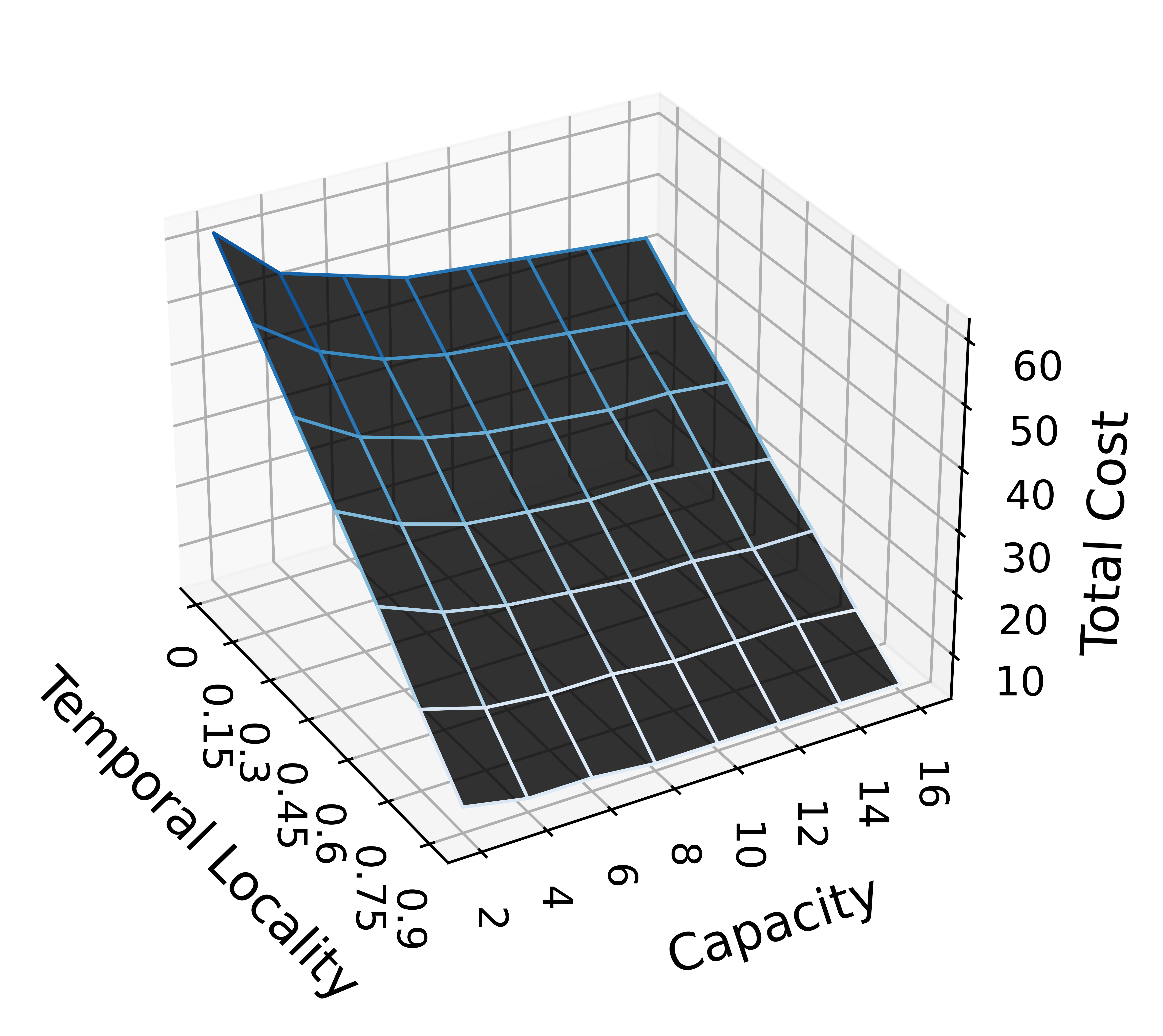}
    \caption{
    }
    \label{subfig: 3d} 
    \end{subfigure}
    \begin{subfigure}[b]{0.33\linewidth}
    \centering
    \includegraphics[width=\textwidth]{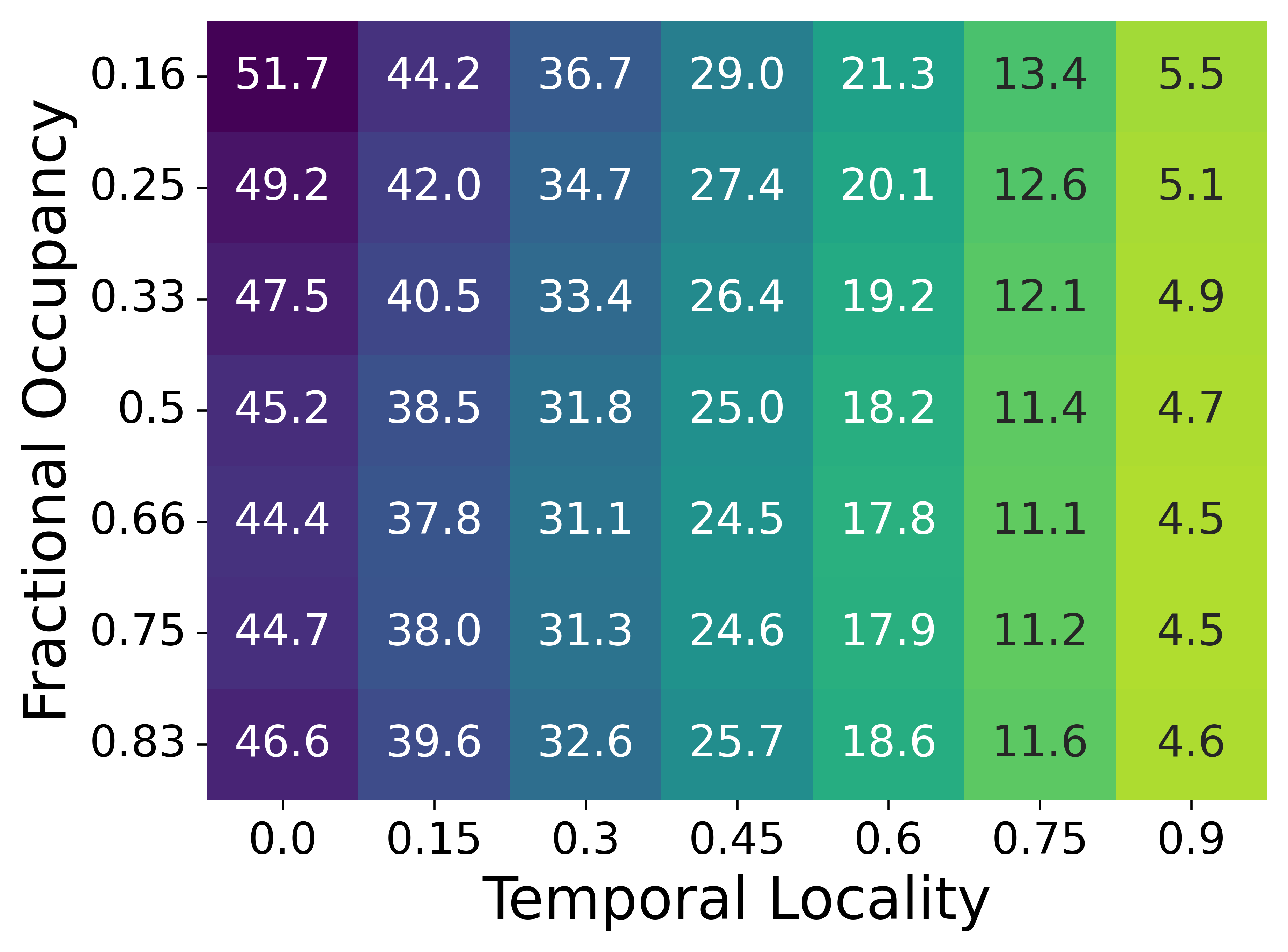}
    \caption{
    }    
    \label{subfig: heatmap} 
    \end{subfigure}
    \caption{ Improvements in the performance of \system by fine-tuning parameters.
    Figures are generated using the~\hyperref[input: temporal]{synthetic dataset} with various locality values.
    (\ref{subfig: toObv}) Comparing the access cost of the \system with fractional occupancy $f=\frac{1}{2}$ to the best possible static algorithm and the demand-oblivious algorithm, all given capacity $c=4$. Access costs are divided by $100$ thousands.
    (\ref{subfig: 3d}) The effect of increasing capacity of \nodes and temporal locality of input on the total cost of the algorithm. The fractional occupancy is set to $f=\frac{1}{2}$ for all capacities. Total costs are divided by $1$ million for this plot.
    (\ref{subfig: heatmap}) Tradeoff between the total cost and the fractional occupancy, given a range of temporal localities. The capacity of nodes is set to $12$. The number in each cell represents the cost, which are divided by $1$ million.
    }
    \label{fig: temporal}  
\end{figure*}

\section{Experimental Evaluation}
\label{sec: evaluation}
We complement our analytical results by evaluating \system on multiple datasets. Concretely, we are interested in answering the following questions:
\begin{enumerate} [start=1,label={ Q\arabic*}]
    \item How does the access cost of our algorithm compare to the statically-optimal algorithm (optimized based on frequencies) and a demand-oblivious algorithm?
    \label{q: zero}
    \item How does additional capacity improve the performance of the online \system, given fixed fractional occupancy of each level?
    \label{q: one}
    \item What is the best initial fractional occupancy for the online \system, given a fixed capacity?
    \label{q: two}
\end{enumerate}
Answers to these questions would help developers tune parameters of the \system based on their requirements and needs. Before going through results, we describe the setup that we used: Our code is written in Python~3.6 and we used seaborn~0.11~\cite{Waskom2021} and Matplotlib~3.5~\cite{Matplotlib} libraries for visualization.
Our programs were executed on a machine with 2x Intel Xeons E5-2697V3 SR1XF with 2.6 GHz, 14 cores each, and a total of 128 GB DDR4 RAM.
\subsection{Input}

\begin{itemize}
    \item  \label{input: facebook} \textbf{Real-world dataset}: Our real-world dataset is communications between servers inside three different Facebook clusters, obtained from~\cite{sigmetrics20complexity}. We post-processed this dataset for single-source communications. Among all possible sources, we chose the most frequent source.
    \item \label{input: temporal} \textbf{Synthetic dataset}: We use the Markovian model discussed in~\cite{sigmetrics20complexity, icdcs22} for generating sequences based on a temporal locality parameter which ranges from $0$ (uniform distribution, no locality) to $0.9$ (high temporal locality). 
    Our synthetic input consists of $65,535$ items and $1$ million requests. 
    For generating such a dataset, we start from a random sample of items. We post-process this sequence, overwriting each request with the previous request with the probability determined by our temporal locality parameter. After that, we execute the second post-processing to ensure that exactly $65,535$ items are in the final trace.
\end{itemize}

\subsection{Algorithm setup}
We use SHA-512~\cite{DobraunigEM16Sha} from the $hashlib$-library as the hash function in our implementation, approximating the 
uniform distribution for generating addresses of items.
In order to store items in a \node we used a linked list, and when we move an item to a \node that is already full with other items, items are stored in a temporary buffer.
We assume starting from a pre-filled tree with items, a tree which respects the fractional occupancy parameter.

In our experiments, we range the capacities ($c$) from $2$ to $16$, and the fractional occupancies ($f$) from $0.16$ to $0.83$. Due to the random nature of our algorithms and input generations, we repeat each experiment up to $100$ times to ensure consistency in our results.

\begin{figure}[t]
    \centering
    \begin{subfigure}[b]{0.49\linewidth}
    \centering
    \includegraphics[width=\textwidth]{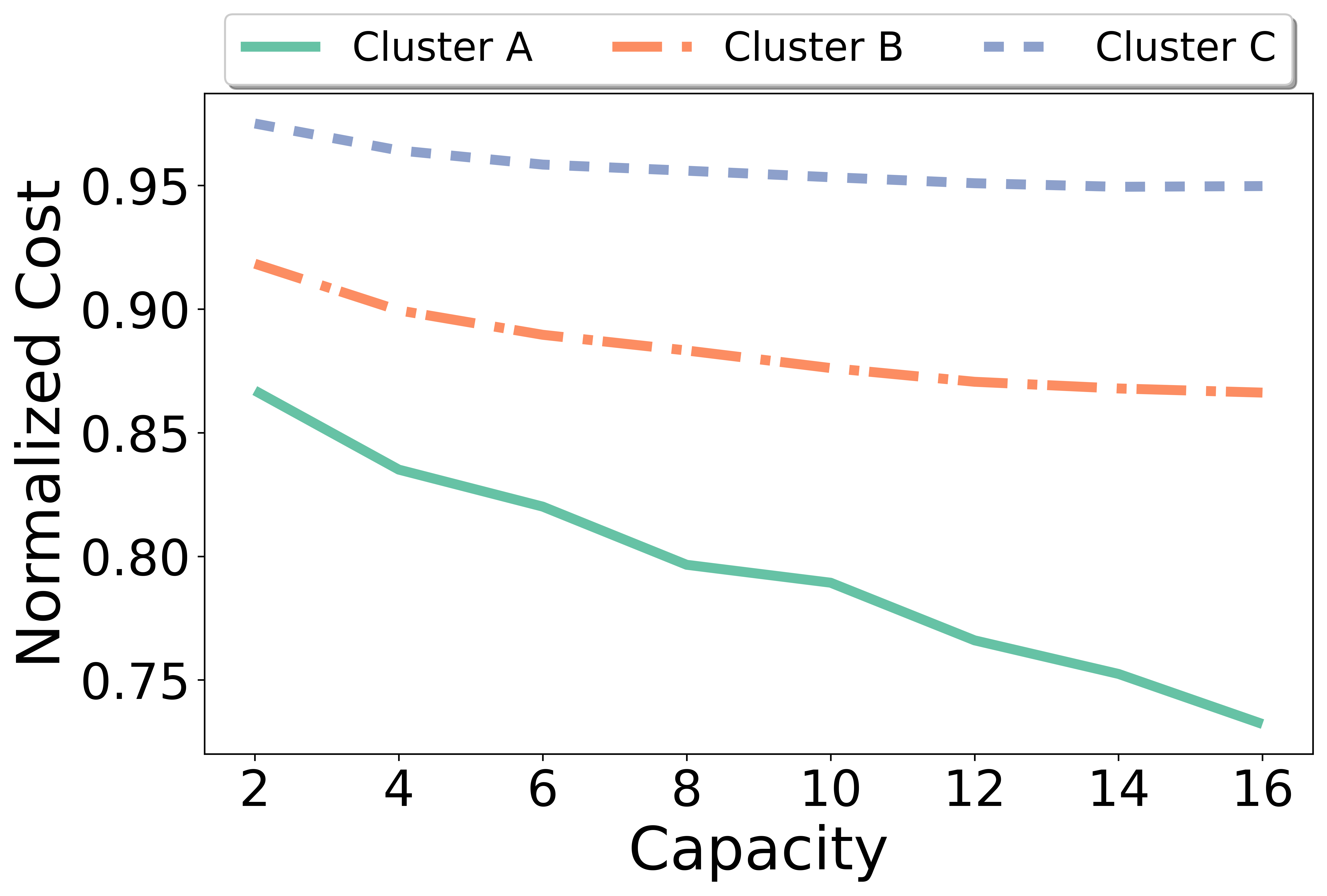}
    \caption{
    }
    \label{subfig: fb-cap}  
    \end{subfigure}
    \begin{subfigure}[b]{0.49\linewidth}
    \centering
    \includegraphics[width=\textwidth]{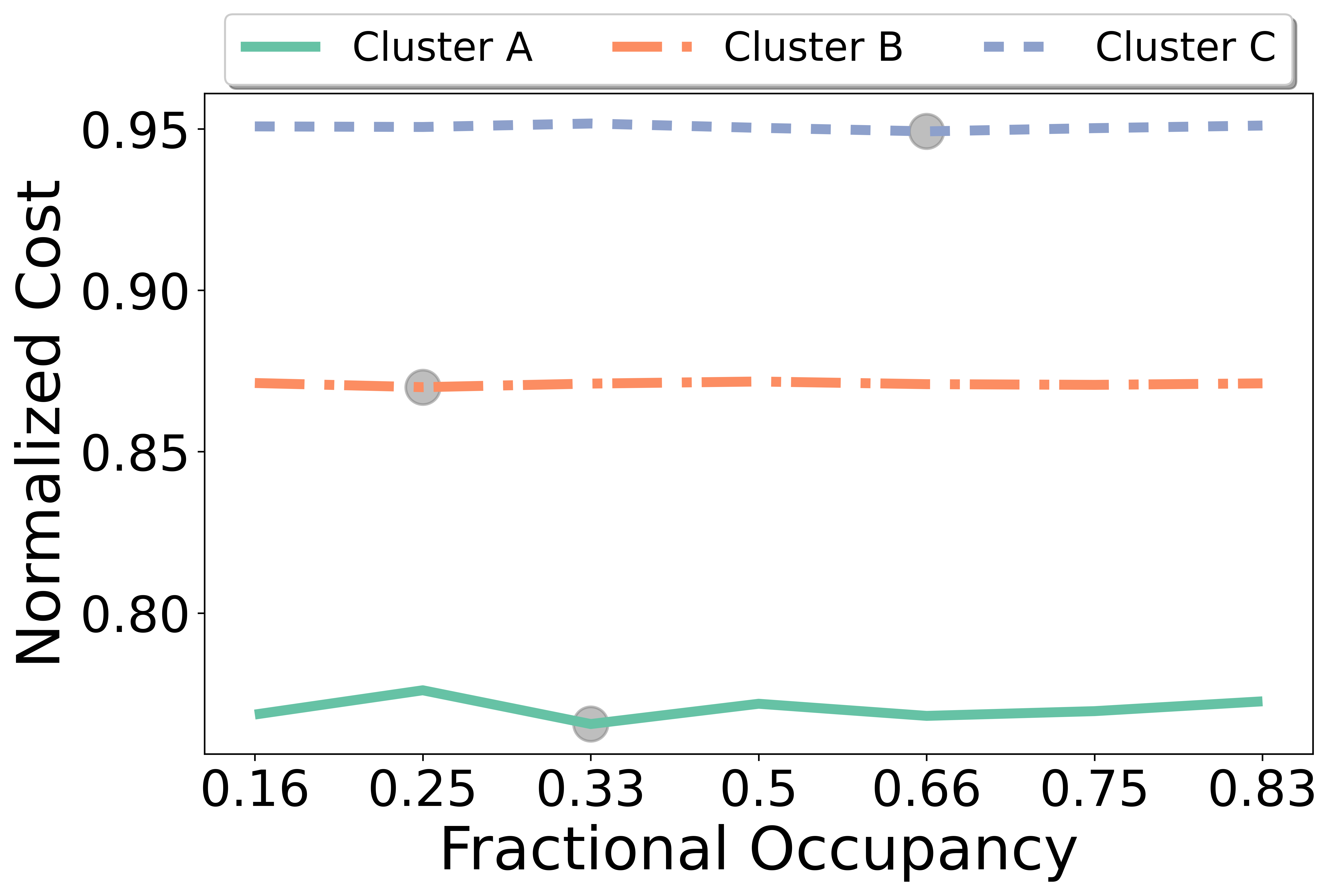}
    \caption{
    }
    \label{subfig: fb-f} 
    \end{subfigure}
    \caption{ Improvements in the normalized access cost of the algorithm by changing \system parameters. These results are obtained based on communications of the most frequent source from three clusters of the ~\hyperref[input: facebook]{real-world dataset}.
    Costs are normalized by the cost of the demand-oblivious algorithm.
    (\ref{subfig: fb-cap}) Changes in the normalized cost by varying capacity. Fractional occupancy is set to $f=\frac{1}{2}$. 
    (\ref{subfig: fb-cap}) Changes in the normalized cost by varying fractional occupancy. Gray dots show the minimum values. Capacity of \nodes is set to $12$. 
    }
    \label{fig: facebook}  
\end{figure}

\subsection{Results}
The performance of \system improves significantly with the increased temporal locality, as can be seen in Figure~\ref{fig: temporal}. Furthermore, we have the following empirical answers to questions proposed at the beginning of this section:

\begin{enumerate} [start=1,label={ A\arabic*}:]
    \item The \system improves the access cost significantly, with increased temporal locality, as shown in Figures~\ref{subfig: toObv}, which compares the access cost of \system to static and demand-oblivious algorithms.
    \label{a: zero}
    \item As the Figures~\ref{subfig: 3d} and~\ref{subfig: fb-cap} show, increasing capacity reduces the cost of the algorithm. However, as we can see, this increase slows down beyond capacity to $8$, and hence this value can be considered as the best option for practical purposes.
    \label{a: one}
    \item As discussed at the end of the \S\ref{sec: online} and can be seen in Figures~\ref{subfig: heatmap} and~\ref{subfig: fb-f}, the lowest cost can be achieved with fractions higher or lower than $\frac{1}{2}$, but $f=\frac{1}{2}$ is near optimal in most scenarios.
    \label{a: two}
\end{enumerate}

\section{Additional Related Work}
\label{sec: related work}

Self-adjusting lists and trees have already been studied intensively in the context of data structures. 
The pioneering work is by Sleator and Tarjan~\cite{ST1985}, who initiated the study of the dynamic list update problems and who also introduced the move-to-front algorithm, inspiring many deterministic~\cite{albers1994competitive,KamaliSurvey2013} and
randomized~\cite{Albers2020,albers1995combined,garefalakis1997new,Reingold1994} approaches for datastructures, as well as other variations of the problem~\cite{albers2008list}.

\begin{table}[t]
\begin{center}
\begin{tabular}{ c | c c c c}
   \hline
   \multicolumn{1}{c|}{Data Structure}  &  \cellcolor{gray!20} Operation
    & \multicolumn{1}{|c|}{Ratio}
    & \cellcolor{gray!20} Search
   \\ \hline
    \cellcolor{gray!20} Splay Tree~\cite{splaytrees}
    & \cellcolor{blue!20} Rotation 
    &
     \cellcolor{red!20} $O(\log n)$&
     \cellcolor{green!20} Yes 
   \\
    Greedy Future~\cite{GreedyBSTFirst}
    & \cellcolor{blue!20} Rotation 
    &
        \cellcolor{red!20} $O(\log n)$&
        \cellcolor{green!20} Yes 
   \\
    \cellcolor{gray!20}  Tango Tree~\cite{TangoTree}
    & \cellcolor{blue!20} Rotation
    &\cellcolor{red!20} $\theta(\log \log n)$& \cellcolor{green!20} Yes 
    \\
    Adaptive Huffman~\cite{CormackH84Huffman}& 
    \cellcolor{blue!20} Subtree swap&
    \cellcolor{green!20} $\theta(1)$& \cellcolor{red!20} No 
   \\
     \cellcolor{gray!20} Push-down Tree~\cite{ton22push}&
    \cellcolor{blue!20} Item swap &
     \cellcolor{green!20} $\theta(1)$&
     \cellcolor{red!20} No 
   \\
   SeedTree&
   \cellcolor{blue!20} Item movement &
   \cellcolor{green!20} $\theta(1)$& 
    \cellcolor{green!20} Yes 
   \\
  \hline
\end{tabular}
\end{center}
\caption{Comparison of properties of self-adjusting tree data structures. The best known competitive ratio (to this date) is in terms of the data structure's respective cost model and optimal offline algorithm. We note that none of the above trees considers additional capacity, except for our model.}
\label{table: tree}
\end{table}
Self-adjusting binary search trees also aim to keep recently used elements close to the root, similarly to our approach in this paper (a summary of results is in Table~\ref{table: tree}). However, adjustments in binary search trees are based on rotations rather than the movement of items between different nodes.
One of the well-known self-adjusting binary search trees is the splay tree~\cite{splaytrees}, although it is still unknown whether this tree is dynamically optimal; the problem is still open also for recent variations such as Zipper Tree~\cite{Zippertree}, Multi Splay Tree~\cite{MultiSplay}
and Chain Splay~\cite{chainSplay} which improve the $O(\log n)$ competitive ratio of the splay tree to $O(\log \log n)$. For Tango Trees~\cite{TangoTree}, a matching $\Omega(\log \log n)$ lower bound is known. We also know that if we allow for free rotations after access, dynamic optimally becomes possible~\cite{search-optimality}.
We also point out that some of these structures, in particular, multi splay tree and chain splay, benefitted from additional memory as well, however, there it is used differently, namely toward saving additional attributes for each node. Another variation which was first proposed by Lucas~\cite{GreedyBSTFirst} in 1988 is called Greedy Future. This tree first received attention as an offline binary search tree algorithm~\cite{GreedyBST1,GreedyBST2}, but then an $O(\log n)$ amortized time in online settings was suggested by Fox~\cite{GreedyBSTFox}. Greedy Future has motivated researchers to take a geometric view of online binary search trees~\cite{GreedyBST2,IaconoSurvey}.
We note that in contrast to binary search trees, our local tree does not require an ordering of the items in the left and right subtrees of a node.

Self-adjusting trees have also been explored in the context of coding, where for example adaptive Huffman coding~\cite{CormackH84Huffman,FGK85Huffman,MilidiuLP99Huffman,moffat2019huffman,vitter1987design}
is used to minimize the depth of most frequent items.
The reconfiguration cost, however, is different: in adaptive Huffman algorithms, two subtrees might be swapped at the cost of one.

A few data structures have tried to achieve a better competitive ratio by expanding and altering binary search trees (see Table~\ref{table: other} for a summary): The first example, PokeTree~\cite{Poketree}, adds extra pointers between the internal nodes of the tree and achieves an $O(\log \log n)$ competitive ratio in comparison to an optimal binary search tree.
There are also self-adjusting data structures based on skip lists~\cite{Pugh90Skip-Main,Avin20Skip-Iosif}, which have been introduced as an alternative for balanced trees that enforce probabilistic balancing instead. A biased version of skip lists was considered in~\cite{BagchiBG05Skip-Biased}, and later on, a statically optimal variation was given in~\cite{CirianiFLM07-Skip-Static-Optimality} and a dynamic optimal version in a restricted model in~\cite{BoseDL08-Skip-Dynamic}.
Another example is Iacono's working set structure~\cite{IaconoWSStructure} which combines a series of self-adjusting balanced binary search trees and deques, achieving  a worst-case running time of $O(\log n)$, however, it lacks the dynamic optimality property.
We are not aware of any work exploring augmentations to improve the competitive ratio of these data structures.

Our work is also motivated by emerging self-adjusting datacenter networks.
Recent optical communication technologies enable datacenters to be reconfigured quickly and frequently~\cite{avin2017demand,icdcs22,infocom19dan,ballani2020sirius,chen2014osa,fleet,ancs18,ghobadi2016projector,spaa21rdcn,opera,rotornet}, see~\cite{sigact19} for a recent survey.
The datacenter application mentioned in our paper is based on the matching model proposed by~\cite{sigmetrics22cerberus}. Recently~\cite{matchingIosif} introduced an online algorithm for constructing self-adjusting networks based on this model, however the authors do not provide dynamic optimality proof for their method.

It has been shown that demand-aware and self-adjusting datacenter networks can be built from individual trees~\cite{avin2018toward}, called ego-trees, which are used in many network designs~\cite{avin2017demand,infocom19dan,apocs21renets,DistributedSelfAdjusting}, and also motivate our model.
However, until now it was an open problem how to design self-adjusting and constant-competitive trees that support local routing and adjustments, a desirable property in dynamic settings. 

Last but not least, our work also features interesting connections to peer-to-peer networks~\cite{ConsistentKarger97,ConsistentChord03}.
It is known that consistent hashing with previously assigned and fixed capacities allows for significantly improved load balancing~\cite{ AamandKT21,MirrokniTZ18}, which has interesting applications and is used, e.g., in Vimeo's streaming service~\cite{rodland2016improving} and in Google's cloud service~\cite{MirrokniTZ18}.
Although these approaches benefit from data structures with capacity, these approaches are not demand-aware.

\section{Conclusion and Future Work}
\label{sec: conclusion}

This paper presented and evaluated a self-adjusting and local tree,  \system, which adapts towards the workload in an online, constant-competitive manner. \system supports a capacity augmentation approach, while providing local routing, which can be useful for other self-adjusting structures and applications as well. 
We showed a transformation of our algorithm into the matching model for application in reconfigurable datacenters, and evaluated our algorithm on synthetic and real-world communication traces. The code used for our experimental evaluation is available at \url{github.com/inet-tub/SeedTree}.

We believe that our work opens several interesting avenues for future research. In particular, while we so far focused on randomized approaches, it would be interesting to explore deterministic variants of  \system. Furthermore, while trees are a fundamental building block toward more complex networks (as they, e.g., arise in datacenters today), it remains to design and evaluate networks based on \system.
\begin{table}[t]
\begin{center}
\begin{tabular}{ c | c  c c }
   \hline
   \multicolumn{1}{c|}{Data Structure} &  \cellcolor{gray!20} Structure
   &  \multicolumn{1}{|c|}{Ratio}
   \\ \hline
    \cellcolor{gray!20}
    Iacono's structure~\cite{IaconoWSStructure}
    & \cellcolor{blue!20} Trees \& deques  
    & \cellcolor{red!20} $O(\log n)$ 
   \\
       Skip List~\cite{Pugh90Skip-Main} 
       & \cellcolor{blue!20} Linked lists 
       &  \cellcolor{red!20} $O(\log n)$
   \\
    \cellcolor{gray!20}  PokeTree~\cite{Poketree}
    & \cellcolor{blue!20} Tree \& dynamic links
    & \cellcolor{red!20} $O(\log \log n)$ 
   \\
    SeedTree
     & \cellcolor{blue!20} Tree 
     & \cellcolor{green!20} $\theta(1)$ 
   \\
  \hline
\end{tabular}
\end{center}
\caption{Comparison with other self-adjusting data structures that support local-search. The best known competitive ratio (to this date) is in terms of the data structure's respective cost model and optimal offline algorithm. We note that none of the other data structures considers capacity in their design.}
\label{table: other}
\end{table}

\bibliography{Greedy}
\end{document}